\documentclass{article}

\begin{document}
\begin{center}
\underline{{\large \bf
\begin{minipage}[h]{\linewidth}
THE ''FORGOTTEN'' PROCESS : \\
the emission stimulated by matter waves.
\end{minipage}}}

\vspace{5mm}

Philippe Tourrenc$^1$, Marie-Christine Angonin$^1$ and Peter Wolf$^{\ 2,3}$
\vspace{8mm}

$^1${\it Universit\'e Pierre et Marie Curie, ERGA, case 142, 4 Pl.
Jussieu, 75252 Paris, CEDEX 05, France} \\
$^2${\it BNM-SYRTE, Observatoire de Paris, 61 Av. de l'Observatoire,
75014 Paris, France} \\
$^3${\it Bureau International des Poids et Mesures, Pavillon de Breteuil,
92312 S\`evres, CEDEX, France}
\end{center}

\vspace{10mm}

{\bf Abstract:} In a famous paper where he introduces the $A$ and $B$
coefficients, Einstein assumed that atomic decays of excited atoms can be
stimulated by light waves. Here we also include atomic decays stimulated by
atomic waves. It is necessary to change the Maxwell-Boltzmann statistics of
thermal equilibrium into Bose-Einstein statistics and to introduce a
coefficient $C$ which complements the list of the coefficients introduced by
Einstein.

Stimulated emission of light can be considered as the first step towards the
laser. Similarly, stimulated production of matter waves can be considered as
the basic phenomenon for an atom-laser.

Most of the results that we obtain here are not new. However, our method is
very close to elementary classical physics and emphasizes the symmetry
between electromagnetic and matter waves from various points of view. The level 
of the discussion is indirect for advanced undergraduate and possible 
direct for graduate.

\section{Introduction}

In 1916 and 1917, Einstein introduced the well known $A$ and $B$
coefficients \cite{Albert}. For the first time the stimulated emission of
photons by electromagnetic radiation was suggested. In the present paper we
generalize the argument of Einstein. We assume that an atom has an internal
structure and that a matter wave describes its space-time behavior. The
decay of an atom into its ground state can happen spontaneously. It can also
be stimulated by the electromagnetic waves or the matter waves which bathe
the atom. The latter (decay stimulated by matter waves) has been "forgotten"
until the 1990s when it appeared in the context of Bose Einstein
Condensation (BEC) and the atom laser \cite{XIIproc,Wiseman,Spreeuw,Borde}.

In his papers, Einstein put forward "a derivation of Planck's formula [...]
closely related to Wien's original argument". Our goal is different: we want
to emphasize the symmetry between electromagnetic and matter waves. However,
we will follow Einstein and derive a simple description of the "forgotten"
process via the introduction of a new coefficient $C$ in addition to the
Einstein coefficients $A$ and $B$.

We first summarize the usual approach describing atom-photon interactions
using Maxwell-Boltzmann statistics and the Einstein $A$ and $B$
coefficients. This is followed by a short summary of matter waves and their
properties that are relevant to our purpose. In section 4 we restore the
symmetry between electromagnetic and matter waves, which we then apply to
the thermal equilibrium (section 5) leading to the derivation of transition
probabilities for all processes. We obtain the usual Einstein coefficients
and a "forgotten" coefficient from these probabilities in section 6 and
conclude in section 7.

\section{The usual approach}

We consider two level atoms where $E_{1}$ and $E_{2}$ are the energy of the
atom in each level (with $E_{2}-E_{1}=\hbar \omega _{0}>0)$. The atoms are
in an ideal electromagnetic cavity with perfectly reflecting walls. 
The thermal equilibrium, at
temperature $T,$ is achieved. 
The case of atoms in free space is obtained when the volume $
V $ of the cavity becomes infinite.

The electromagnetic energy is distributed among the different modes of the
cavity. It is described by Planck's energy spectral density.
\begin{equation}
\rho _{\varphi }\left( \omega ,T\right) =d_{\varphi }\times \frac{4\pi \nu
^{2}}{c^{3}}\times \hbar \omega \times \frac{1}{e^{\hbar \omega /k_{B}T}-1}
\label{rhonu}
\end{equation}
where $k_{B}$ is Boltzmann's constant. \vspace{5mm} Each term in $\rho
_{\varphi }\left( \omega ,T\right) $ has a precise meaning.

$\bullet $ The electromagnetic wave is characterized by a wave vector $
\overrightarrow{k}$. We define $2\pi \,\nu =c\left\Vert \overrightarrow{k}
\right\Vert .$ The electromagnetic waves in an ideal cavity must satisfy
boundary conditions. As a consequence, $\nu $ can take only resonant values:
the number of possible values between $\nu $ and $\nu +d\nu $ is $d{\mathcal{
N}}_{\nu }=\frac{4\pi \nu ^{2}}{c^{3}}d\nu \times V$ where $c$ is the speed
of light in vacuum and $V$ the volume of the cavity.

$\bullet $ The energy of a photon is $\hbar \omega $, where $\omega $ is the
angular frequency. For an electromagnetic field the relation $\omega =2\pi
\nu $ holds true.

$\bullet $ Once $\overrightarrow{k}$ is given, $\omega $ is known. There
remain however $d_{\varphi }=2$ polarizations. The first factor, $d_{\varphi
}$ in relation (\ref{rhonu}), is precisely the number of polarizations (the
degeneracy of photonic states) of the electromagnetic waves. Therefore, in
an ideal cavity, the number of modes in the interval $\left[ \nu ,\,\nu
+d\nu \right] $ is
\begin{equation}
d{\mathcal{N}}_{m}=d_{\varphi }\times \frac{4\pi \nu ^{2}}{c^{3}}d\nu \times
V  \label{dnmnu}
\end{equation}

$\bullet $ Finally, using Bose-Einstein statistics, the mean number of
photons in a mode at thermal equilibrium is $\frac{1}{e^{\hbar \omega
/k_{B}T}-1}$.

The spectral energy density (\ref{rhonu}) represents the energy density per
interval $d\nu$. It is sometimes useful to define a quantity $\tilde{\rho}
_\varphi(\omega ,T)$ that represents the energy density per corresponding
interval $d\omega$, i.e.

\begin{equation}
\tilde{\rho}_\varphi(\omega ,T) = \rho_\varphi(\omega ,T)\frac{d\nu}{d\omega}
.  \label{rhorho}
\end{equation}
For electromagnetic waves we have $\tilde{\rho}_\varphi(\omega ,T) =
\rho_\varphi(\omega ,T)/2\pi$. \vspace{5mm}

Thermal equilibrium in the cavity is secured by a connection to a thermal
reservoir at temperature $T.$ We assume the following properties:

{\bf A1} First we assume that the reservoir can exchange some energy with
the atoms in order that the Maxwell-Boltzmann equilibrium law is fulfilled:
\begin{equation}
\frac{N_{2}}{N_{1}}=\frac{d_{2}}{d_{1}}e^{-\hbar \omega _{0}/k_{B}T}
\label{NbNa}
\end{equation}
where $N_{1}$ (or $N_{2}$) is the number of atoms with energy $E _{1}$ (or $
E_{2}$) and $d_{1}$ (or $d_{2}$) is the degeneracy of level $E_{1}$ (or $
E_{2}$).

{\bf A2} Second we assume that photons can be either absorbed or emitted
by the reservoir to achieve the equilibrium spectral density, $\rho
_{\varphi }\left( \omega ,T\right) ,$ above.

{\bf A3} Finally we assume that once the thermal equilibrium is achieved,
it remains without the help of the reservoir, although it is a statistical
equilibrium where absorption and emission of photons, and excitation and
decay of atoms, continue to happen.

\vspace{5mm}

Assuming conservation of energy, two kinds of effects are possible:

{\bf 1. Spontaneous mechanisms} happen even when no radiation is present
in the cavity. This is the case of a spontaneous decay of an atom from
energy $E_{2}$ to energy $E_{1}$ with the emission of a photon whose angular
frequency is $\omega _{0}$. The number of spontaneous decays per unit time
is $A\times N_{2}$. In a large enough cavity two neighboring modes have very
similar properties and their frequencies can be considered to belong to a
continuum. In such a case, $A$ is a characteristic of the atom alone,
independent of the cavity dimensions. We do not consider spontaneous
excitation where energy would be created from nothing.

{\bf 2. Stimulated mechanisms} happen only because some radiation is
already present. Two mechanisms are actually possible: i) the absorption of
a photon with energy $\hbar \omega _{0}$ by an atom which is therefore
excited from energy $E_{1}$ to energy $E_{2}$ and ii) the emission of a
photon with energy $\hbar \omega_{0}$ by an atom which decays from energy $
E_{2}$ to energy $E_{1}$. The number of events per unit time of each
mechanism is $B_{\left( abs\right) }\times \rho _{\varphi }\left( \omega
_{0},T\right) N_{1}$ and $B_{(em)}\times \rho _{\varphi }\left( \omega
_{0},T\right) N_{2}$, where $\omega _{0}=2\pi \nu _{0}$. \vspace{5mm}

The equilibrium condition is
\begin{equation}
A N_{2}+B_{(em)}\rho _{\varphi }\left(\omega
_{0},T\right)N_{2}=B_{(abs)}\rho _{\varphi }\left( \omega _{0},T\right)N_{1}.
\label{stat}
\end{equation}

Within the preceding theoretical framework, one finds that the thermal
equilibrium (Eq.~(\ref{stat}) with Eqs.~(\ref{rhonu}) and (\ref{NbNa})) is
possible at any temperature $T$ if and only if
\begin{equation}
B_{(abs)}=\frac{d_{2}}{d_{1}} B_{(em)} \hspace{4mm} \mathrm{and} \hspace{4mm}
A=\frac{8\pi \nu _{0}^{2}}{c^{3}}\hbar \omega _{0}B_{(em)}
\end{equation}

Therefore the existence of spontaneous emission ({\it i.e.} $A\neq 0)$,
implies that absorption and stimulated emission also exist ({\it i.e.} $
A\neq 0 \Rightarrow B_{(em)}\neq 0$ and $B_{(abs)}\neq 0)$.

These results are well known and can be found in many textbooks \cite
{Yariv,Sargent,Hilborn}.

\section{Matter waves}

Matter waves were introduced in de Broglie's thesis in 1924. At this time
the photoelectric effect had already been interpreted (Einstein 1905) and
the Compton effect observed (Compton 1923). Thus, it was already known that
light could behave as a flow of particles called photons \cite{photon}. The merit of Louis de Broglie idea was to reverse the
proposition and claim that particles such as electrons could behave like
waves.

Since these early days, the similarity between light waves and matter waves
never ceased to be emphasized, from the first observation of electronic
interferences by Davisson and Germer in 1927 \cite{Davisson} 
to atom interferometry, and
recently Bose-Einstein condensation of atoms, 
first achieved in 1995 \cite{Anderson}.

Several examples can be given of matter waves and their applications: the
electronic microscope and the Collela-Overhauser-Werner experiment with
neutrons \cite{COW} are well known. More recently laser cooling of atoms has
lead to the development of atom interferometry used, for example, for ultra
precise inertial sensors such as gravimeters \cite{Peters}, gradiometers
\cite{Snadden} and gyroscopes \cite{Gustavson}. Ultimately the use of the
forgotten process to produce Bose-Einstein condensates, i.e. coherent beams
of matter waves with all atoms in the same mode, should result in atom
lasers based on amplification of the matter fields via the forgotten
process. These are expected to lead to significant improvements in matter
wave interferometry and to new applications. Presently, what is often called
an atom laser is a Bose-Einstein condensate which has been produced by other
means.

Despite its complex structure, an atom can be considered as a particle with
a mass $M$ depending on its internal energy : $M=M_{0}+E/c^{2}$ where $E$ is
the internal energy and $M_{0}$ the mass of the atom in its fundamental
level. The total energy of the atom is then

\begin{equation}
\hbar \omega =\sqrt{\left(M c^{2}\right)^{2}+\left( c\hbar \overrightarrow{k}
\right)^{2}} \approx M c^2 + \frac{\hbar^2 \overrightarrow{k}^2}{2 M}
\label{Eat}
\end{equation}
where $\omega$ is its angular frequency, $\hbar \overrightarrow{k}$ its
momentum and the non-relativistic approximation holds for $\hbar \Vert
\overrightarrow{k}\Vert /M \ll c$.

An atom is generally characterized by its angular momentum, $\overrightarrow{
J}$, which plays the role of an intrinsic spin. The value of $
\overrightarrow{J}^{2}$ is $\hbar ^{2}j\left(j+1\right)$. General results in
quantum mechanics lead to the conclusion that $2j$ is an integer which
depends on the internal state of the atom, and that the atom is a boson when
$j$ itself is an integer. This is the only case that we consider \cite{sensitivity}.

The energy can be degenerate. Therefore, for a given mass, the internal
state is a multi-component vector which belongs to a $d_a$ dimensional space.

Keeping the definition $2\pi\,\nu=c\left\Vert\overrightarrow{k}\right\Vert$,
the fundamental difference with respect to photons is that now the relation
between $\omega $ and $\nu $ is the dispersion relation obtained from Eq.~(
\ref{Eat})
\begin{equation}
\hbar \omega =\sqrt{\left( M c^{2}\right) ^{2}+\left( 2\pi \hbar \nu \right)
^{2}}.  \label{disp}
\end{equation}

Let us imagine that matter waves can be trapped in an ideal cavity where
they appear as a superposition of the resonant modes of the cavity. The
number of resonance values of $\nu$ in the interval $\left[ \nu ,\nu +d\nu 
\right]$ is

\begin{equation}
d{\mathcal{N}}_{\nu }=\frac{4\pi \nu ^{2}}{c^{3}}d\nu \times V  \label{dN}
\end{equation}
and the number of modes $d{\mathcal{N}}_{m}=d_a \times d{\mathcal{N}}_{\nu}$.

As we consider bosons only, the mean number of particles per mode at thermal
equilibrium is given by Bose-Einstein statistics
\[
\overline{n}=\frac{1}{e^{\left( \hbar \omega -\mu \right) /\left(
k_{B}T\right) }-1}
\]
where $\mu $ is the chemical potential (see ref. \cite{Landau} for instance).

In section 2, we assumed that the atoms are not able to be created or
annihilated, only their internal energy could change by emission or
absorption of photons. However, considering what we know about the origin of
the Universe, we have to admit that atoms can be created too. Of course, the
corresponding mechanism can be complicated and slow but annihilation and
creation of atoms via exchange of energy with the thermal reservoir
 are possible as well as annihilation and creation of
photons; this is a matter of principle not of order of magnitude. Thus the
number of atoms in the cavity must be determined from the condition of
thermal equilibrium. The same happens to the photons with the same
consequence, {\it i.e.} $\mu =0\ $(see the black body radiation in 
\cite{Landau}). Therefore :
\begin{equation}
\overline{n}=\frac{1}{e^{\left( \hbar \omega \right) /\left( k_{B}T\right)
}-1}.  \label{nth}
\end{equation}

Therefore, the spectral energy density is
\begin{equation}
\rho _{a}\left( \omega ,T\right) = d_a \times \frac{4\pi \nu ^{2}}{c^{3}}
\times \hbar \omega \times \frac{1}{e^{\left( \hbar \omega \right) /\left(
k_{B}T\right) }-1}.  \label{spect}
\end{equation}

Analogously to the electromagnetic case one can introduce a corresponding
energy density per interval of angular frequency $\tilde{\rho}_{a}\left(
\omega ,T\right)$ defined by Eq.~(\ref{rhorho}). But now $d\nu /d\omega$ is
obtained from the dispersion relation (\ref{disp}) and therefore

\begin{equation}
\tilde{\rho}_{a}\left( \omega ,T\right) = \frac{\rho _{a}\left( \omega
,T\right)}{2\pi} \frac{\hbar\omega}{\sqrt{\left(\hbar\omega\right)^2 -
\left(Mc^2\right)^2}}= \frac{\rho _{a}\left( \omega ,T\right)}{2\pi}\frac{
\omega}{2\pi\nu} .  \label{tilderhoa}
\end{equation}

One can check that the relations (\ref{disp}) to (\ref{tilderhoa}) are the
same for matter waves and for light waves when setting $M=0$ and $
d_a=d_\varphi =2$ for the photon.

\section{Restoring the symmetry between electromagnetic and matter waves}

In section 2, the atoms and the photons have been considered from two very
different points of view. For instance, the decay of an atom could be
stimulated by the presence of the photons but the presence of the atoms did
not produce any similar effect.

First, in order to restore the similarity between the atoms and the photons,
let us modify the notation. Now, $a$ is an atom in its fundamental state of
energy $E_{a}$. This atom can absorb a photon $\varphi $, the corresponding
excited state is $a\varphi $ whose internal energy is $E_{a\varphi }$. We
assume that thermal equilibrium is driven by the chemical-like equation
\begin{equation}
a\varphi \rightleftharpoons a+\varphi  \label{reac}
\end{equation}
where $\varphi $, $a$ and $a\varphi $ are bosons described by waves (i.e.
electromagnetic waves and matter waves) which are trapped in an ideal cavity
at temperature $T$.

An atom is considered as the quantum associated to a matter wave, therefore $
N_{a\varphi}$ and $N_{a}$ are now occupation numbers of matter-wave-modes,
similar to the occupation number of the electromagnetic modes (i.e. the
number of photons).

The mass of the atom depends on its internal energy. Thus we can interpret
the change of the internal energy as the annihilation of an atom with the
initial value of the mass and the creation of an atom with the final value
of the mass. \vspace{5mm}

The atoms and the photons are assumed to be trapped in an ideal cavity
without losses. The cavity is coupled to a thermal reservoir at temperature $
T$. We assume the following properties:

{\bf B1} The photons and the atoms occupy respectively
electromagnetic-modes and matter-wave-modes. The equilibrium is achieved
when the mean number of quanta per mode is given by the Bose-Einstein
statistics $\overline{n}=\frac{1}{e^{\hbar \omega /k_{B}T}-1}\ $i.e. we
substitute Eq.~(\ref{nth}) into Eq.~(\ref{NbNa}).

{\bf B2} Second we assume that photons and atoms can be either absorbed
or emitted by the reservoir, in order to achieve the equilibrium spectral
densities (\ref{rhonu}) and (\ref{spect}).

{\bf B3} Finally we assume that once thermal equilibrium is achieved, it
remains without the help of the reservoir although it is in statistical
equilibrium where annihilation and creation of photons and atoms, continue
to happen.

From the old point of view the dissymmetry between matter and light lies in
the difference between the assumptions {\bf A1} and {\bf A2} of
section 2. Here, this dissymmetry has disappeared but the Maxwell-Boltzmann
statistics used in assumption {\bf A1} has been changed into
Bose-Einstein relativistic statistics. \vspace{5mm}

The various modes of the excited atoms $a\varphi$ are labelled by a set of
indexes called $m$; the modes of the atoms $a$ are labelled by $n$ and the
modes of the photons $\varphi$ by a set of indexes $k$. We use the notation $
\left( a\varphi \right) _{m}$ for an excited atom $a\varphi $ in the mode $m$
, and the similar notations $\left( a\right) _{n}$ and $\left( \varphi
\right) _{k}$. With this notation Eq.~(\ref{reac}) becomes
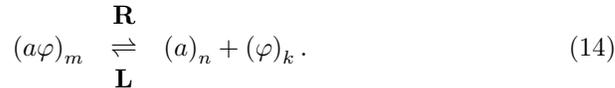
\begin{equation}
\begin{array}{ccc}
& {\bf R} &  \\
\left( a\varphi \right) _{m} & \rightleftharpoons & \left( a\right)
_{n}+\left( \varphi \right)_{k}. \\
& {\bf L} &
\end{array}
\label{reac2}
\end{equation}

Following assumption {\bf B3} above, we assume that the equilibrium is
achieved when, during a given arbitrary time, the number of reactions to the
right (reaction {\bf R} is the same as the number of reactions to
the left (reaction {\bf L}). Moreover we accept the usual
assumption that the number of reactions per unit time is proportional to the
number of quanta in the modes involved.

We introduce the concentrations $\left[ a\varphi \right] _{m}$, $\left[ a
\right] _{n}$ and $\left[ \varphi \right] _{k}$ i.e. the number of atoms or
photons per unit volume respectively in mode $m$, $n$ and $k$. The volume of
the cavity is $V$. We define $\alpha$, $\beta_{(abs)}$, $\beta_{(em)}$ and $
\gamma$:

\noindent (i) The number of spontaneous reactions {\bf R} per unit
time is $\alpha \times \left[ a\varphi \right] _{m}V$.

\noindent (ii) The number of reactions {\bf L} per unit time is $
\beta_{(abs)} \times \left[ \varphi \right] _{k}V\times \left[ a \right]
_{n}V $.

\noindent (iii) Given an excited atom, $a\varphi$, in the mode $m$, we
consider its decay (reaction {\bf L}) stimulated by the presence of the photons
in the mode $k$. The number of such reactions per unit time is $\beta_{(em)}
\times \left[ \varphi \right] _{k}V\times \left[ a\varphi \right] _{m}V$.

With Maxwell-Boltzmann statistics where $M_{0}c^{2}\gg\hbar \omega _{0}$ and
$Mc^{2}\gg k_{B}T$, and within the "broad band" approximation \cite{broad},
 these mechanisms result in Eq.~(\ref
{stat}).

\noindent (iv) Finally, we assume that the reaction {\bf R} can
also be stimulated by the presence of the atoms $a$ in the mode $n$, which
restores the symmetry between electromagnetic waves and matter waves. We
call this mechanism the "forgotten" process \cite{chronology}.
 The number of such reactions per unit
time is $\gamma \times \left[ a\right] _{n}V\times \left[ a\varphi \right]
_{m}V$.

\section{Thermal equilibrium and its consequences}

Compared to the year 1917 the conception of matter has changed dramatically.
We will now explain why this conceptual change leaves the description of
thermal equilibrium practically unchanged. However, we will emphasize the
importance of the "forgotten" process.

Now we assume that equilibrium is achieved when the mean number of quanta
per mode is given by Bose-Einstein statistics and when the number of
reactions {\bf R} per unit time is equal to the number of
reactions {\bf L}. Therefore

\begin{eqnarray}
\left[ a\right] _{n}V&=&\frac{1}{e^{\left( \hbar \omega _{a}\right) /\left(
k_{B}T\right) }-1},  \nonumber \\
\left[ \varphi \right] _{k}V&=&\frac{1}{e^{\left( \hbar \omega _{\varphi
}\right) /\left( k_{B}T\right) }-1},  \label{equilat} \\
\left[a\varphi \right] _{m}V&=&\frac{1}{e^{\left( \hbar \omega _{a\varphi
}\right) /\left( k_{B}T\right) }-1},  \nonumber
\end{eqnarray}

\noindent and
\begin{eqnarray}
\beta_{(abs)}\times \left[ \varphi \right] _{k}V\times \left[ a\right] _{n}V
&=&\alpha \times \left[ a\varphi \right] _{m}V+\beta_{(em)}\times \left[
\varphi \right] _{k}V\times \left[ a\varphi \right] _{m}V  \nonumber \\
&&+\gamma\times \left[ a\right] _{n}V\times \left[ a\varphi \right] _{m}V.
\label{equil}
\end{eqnarray}

The energy of the atom $a$ in the mode $n$ is $\hbar \omega _{a}$, similarly
$\hbar \omega _{\varphi }$ is the energy of the photon in the mode $k$ and $
\hbar \omega _{a\varphi }$ is the energy of the excited atom $a\varphi$ in
the mode $m$. We assume that the energy ($\hbar \omega _{a}$, $\hbar \omega
_{\varphi }$ or $\hbar \omega _{a\varphi })$ defines the mode except for the
polarization of the light and the degeneracy of the internal energy of the
atoms \cite{rectangular}.

One can check that the equality (\ref{equil}) holds true at any temperature
if and only if
\begin{equation}
\omega _{a\varphi }=\omega _{a}+\omega _{\varphi }\hspace{5mm}\mathrm{and}
\hspace{5mm}\beta _{(abs)}=\alpha =\beta _{(em)}=\gamma  \label{zow}
\end{equation}
The first relation in (\ref{zow}) expresses the conservation of energy while
the other relations imply that the "forgotten" process does exist ({\it 
i.e.} $\gamma \neq 0)$ because spontaneous decay is observed (because $\alpha
\neq 0)$.

Let us now verify that at thermal equilibrium the results above are
practically the usual ones when the "forgotten" process is neglected. We
consider Eq.~(\ref{equil}) with $\beta_{(abs)}=\alpha =\beta_{(em)}=\gamma$.
Then the sum of the last two terms is $\gamma\times\left[ \varphi \right]
_{k}V\times \left[ a\varphi \right] _{m}V\left( 1+\frac{\left[ a\right] _{n}
}{\left[ \varphi \right] _{k}}\right) .$ Using Eq.~(\ref{equilat}) it is
easy to check that $\frac{\left[ a\right] _{n}}{\left[ \varphi \right] _{k}}
\ll 1$ even with extreme (impossible!) values of $\hbar \omega _{\varphi }$
and $k_{B}T$. Therefore the contribution $\gamma\times\left[ a\right]
_{n}V\times \left[ a\varphi \right] _{m}V$ in Eq.~(\ref{equil}) is
completely negligible and the "forgotten" process does not play any
significant role at thermal equilibrium.

Moreover one can easily calculate $N_{a\varphi }/N_{a}$ where $N_{a\varphi }$
(respectively $N_{a}$) is the mean number of atoms with energy $E_{a\varphi }=\hbar
\omega _{a\varphi }$ (respectively $E_{a}=\hbar \omega _{a}$). It is the number of
atoms in one mode with energy $\hbar \omega _{a\varphi }$ (respectively $\hbar
\omega _{a}$) times the number of modes with such an energy. From the
preceeding assumptions we obtain

\begin{equation}
N_{a\varphi}=\frac{d_{a\varphi }}{e^{\left( \hbar \omega _{a\varphi }\right)
/\left( k_{B}T\right) }-1} \hspace{5mm} \mathrm{and \ similarly} \hspace{5mm}
N_{a}=\frac{d_{a}}{e^{\left( \hbar \omega _{a}\right) /\left( k_{B}T\right)
}-1}.
\end{equation}
Finally, using the relation $\hbar \omega _{a\varphi }\sim \hbar \omega
_{a}\sim M_{0}c^{2}\gg k_{B}T$ we obtain

\begin{equation}
\frac{N_{a\varphi}}{N_{a}} \simeq \frac{d_{a\varphi }}{d_{a}}\frac{e^{\left(
\hbar \omega _{a}\right) /\left( k_{B}T\right) }}{e^{\left( \hbar \omega
_{a\varphi }\right) /\left( k_{B}T\right) }}=\frac{d_{a\varphi }}{d_{a}}
e^{-\left( \hbar \omega _{a\varphi }-\hbar \omega _{a}\right) /\left(
k_{B}T\right) }
\end{equation}
which is identical to Eq.~(\ref{NbNa}) under the simple change of notation ($
a\varphi \rightarrow 2$, $a \rightarrow 1$ and $\hbar \omega _{a\varphi
}-\hbar \omega _{a}=\hbar \omega _{0})$.

We notice that at thermal equilibrium nothing is significantly modified if
we use Maxwell-Boltzmann statistics instead of Bose-Einstein statistics and
if we neglect the "forgotten" process. However, far from equilibrium this is
not necessarily the case.

Now we can use standard methods to give an estimation of $\gamma$ in the
simplest case of a homogeneous line width.

Let $\alpha$ be the probability per unit time that an excited atom $a\varphi
$, in a given mode $m$ with given energy $\hbar \omega _{a\varphi }$, decays
spontaneously into $a+\varphi$ where the photon $\varphi $ is in the mode $k$
with energy $\hbar \omega _{\varphi }$ and the atom $a$ in the mode $n$ with
energy $\hbar \omega _{a}$. Let us define the probability per unit time, $dp$
, that an excited atom decays spontaneously into a photon with angular
frequency $\omega _{\varphi }\in \left[ \omega_{\varphi k} ,\omega_{\varphi
k} +d\omega \right]$ and a ground state atom with angular frequency $\omega
_{a}\in \left[\omega_{an} -d\omega,\omega_{an} \right]$ where $\omega
_{a\varphi }=\omega _{a}+\omega _{\varphi }$. We then have

\begin{equation}
dp = \alpha \times d_\varphi d_a \times d{\mathcal{N}}_{a+\varphi}
\label{dp1}
\end{equation}
where $d{\mathcal{N}}_{a+\varphi}$ is the number of pairs of resonance
values ($\nu_{a}$, $\nu_{\varphi}$) that satisfy $\omega _{\varphi }\in 
\left[ \omega_{\varphi k} ,\omega_{\varphi k} +d\omega \right]$, $\omega
_{a}\in \left[ \omega_{an}-d\omega ,\omega_{an} \right]$ and $\omega
_{a\varphi }=\omega _{a}+\omega _{\varphi }$.

For the photons, the number of resonant values of $\nu_{\varphi}$, $d{
\mathcal{N}}_\varphi$, over the bandwidth $d\omega$ are given (c.f. Eq.~(\ref
{rhonu})) by the number of possible values of $\nu$, i.e. $\frac{4 \pi \nu^2
}{c^3}d\nu V$, and $2\pi \nu = \omega$ which holds true for photons,
therefore
\begin{equation}
d{\mathcal{N}}_\varphi = \frac{\omega_\varphi \nu_\varphi d\omega}{\pi c^3}V.
\label{dNphi}
\end{equation}
Similarly for the ground state atom, the number of resonant values of $
\nu_{a}$ over the bandwidth $d\omega$ are given by $\frac{4 \pi \nu^2}{c^3}
d\nu V$, but with the relation between $\nu$ and $\omega$ given in Eq.~(\ref
{disp}). This leads to
\begin{eqnarray}
d{\mathcal{N}}_a &=& \frac{\omega_a \nu_a d\omega}{\pi c^3}V =\frac{\omega_a
\sqrt{\left(\hbar\omega_a\right)^2 - \left(M_0 c^2 \right)^2}d\omega}{2
\pi^2 c^3 \hbar}V  \nonumber \\
&\simeq& \frac{\left(M_0 c^2 \right) \sqrt{2 M_0 c^2 E_{k}} d\omega}{2 \pi^2
c^3 \hbar^2}V  \label{dNa}
\end{eqnarray}
where $E_k=\hbar\omega_a-M_0c^2$ is the kinetic energy of $a$, and where we
have used $\hbar\omega_a+M_0c^2 \simeq 2M_0c^2$.

Comparing Eq.~(\ref{dNphi}) to Eq.~(\ref{dNa}) we note that even in extreme
conditions $d{\mathcal{N}}_a$ is much bigger than $d{\mathcal{N}}_\varphi$.
For example, with $E_k\,\approx\,k_BT \approx 10^{-13}$ eV, $M_0c^2 \approx
1 $ GeV and $\hbar\omega_\varphi \approx 20\ $eV we have $d{\mathcal{N}}_a
\approx 10^4 d{\mathcal{N}}_\varphi$. As a result the number of possible
energy pairs $d{\mathcal{N}}_{a+\varphi}$ is entirely determined by $d{
\mathcal{N}}_\varphi$ because for each $\omega_\varphi$ there exists a $
\omega_a$ such that $\omega _{a\varphi }=\omega _{a}+\omega _{\varphi }$
(but not vice-versa), so we have $d{\mathcal{N}}_{a+\varphi} \simeq d{
\mathcal{N}}_\varphi$. Under these conditions Eq.~(\ref{dp1}) becomes
\begin{equation}
dp = \alpha \times d_\varphi d_a \times \frac{\omega_\varphi \nu_\varphi
d\omega}{\pi c^3}V.  \label{dp2}
\end{equation}
On the other hand, under the assumption that the atom is at rest and the
assumption of isotropy, the spontaneous emission of a photon is
characterized by a function $f_\varphi\left(\omega_\varphi - \omega_{\varphi
0} \right)$ which defines the line shape of the emitted light. The quantity $
f_\varphi\left(\omega_\varphi - \omega_{\varphi 0} \right)d\omega$ is
interpreted as the probability that the emitted photon displays an angular
frequency $\omega_\varphi \in \left[ \omega_{\varphi k},\omega_{\varphi k}
+d\omega\right]$. Therefore

\begin{equation}
dp=\frac{1}{t_{sp}}f_\varphi\left( \omega_\varphi - \omega_{\varphi 0}
\right)  \label{dp3}
\end{equation}
where $t_{sp}$ is the time constant which characterizes the spontaneous
emission (i.e. $1/t_{sp}=A$, the Einstein coefficient).

It is then straightforward to obtain $\alpha$ by eliminating $dp$ between
Eqs.~(\ref{dp2}) and (\ref{dp3})

\begin{equation}
\alpha=\beta_{abs}=\beta_{em}=\gamma=\frac{\pi c^3}{d_\varphi d_a
\omega_\varphi \nu_\varphi}\frac{1}{t_{sp}}f_\varphi(\omega_\varphi -
\omega_{\varphi 0})\times \frac{1}{V}.  \label{alpha}
\end{equation}
One can now calculate the probability per unit time, $W_{(em)}$, of
stimulated decay of an excited atom $a\varphi$ in mode $m$ into a photon $
\varphi$ in mode $k$ and a ground state atom $a$ in any mode. First we
notice that the decay due to the photons in mode $k$, towards a special
given mode of $a$, has a probability per unit time $\beta_{(em)} \times 
\left[ \varphi \right] _{k}V$ (see the property (iii) of section 4 above).
Therefore, the probability of a decay towards the various modes $n$ of $a$
which display the same energy is
\begin{equation}
W_{\left( em\right) }=d_{a}\times \beta_{(em)} \times \left[ \varphi \right]
_{k}V = \frac{\pi c^3}{d_\varphi \omega_\varphi \nu_\varphi}\frac{1}{t_{sp}}
f_\varphi(\omega_\varphi - \omega_{\varphi 0}) \times \frac{
u_k(\omega_\varphi)}{\hbar \omega_\varphi}  \label{Wem}
\end{equation}
where we have introduced the electromagnetic energy density of the mode $k$
defined as $u_k(\omega_\varphi)=\left[ \varphi \right] _{k}\times \hbar
\omega_\varphi$.

Similarly the probability per unit time, $W_{(abs)}$, of absorption of a
photon $\varphi$ in mode $k$ by a ground state atom $a$ in mode $n$ to form
an excited atom $a\varphi$ in any mode is
\begin{equation}
W_{\left( abs\right) }=d_{a\varphi}\times \beta_{(em)} \times \left[ \varphi 
\right] _{k}V =\frac{d_{a\varphi}}{d_a}\frac{\pi c^3}{d_\varphi
\omega_\varphi \nu_\varphi}\frac{1}{t_{sp}}f_\varphi(\omega_\varphi -
\omega_{\varphi 0}) \times \frac{u_k(\omega_\varphi)}{\hbar \omega_\varphi}.
\label{Wabs}
\end{equation}
Expressions (\ref{Wem}) and (\ref{Wabs}) are well known in the usual laser
theory and can be found in many textbooks (e.g. \cite{Yariv}).

On the other hand, the "forgotten" process leads to new results. We
calculate the probability per unit time, $W_f$, that an excited atom $
a\varphi$ in mode $m$ undertakes the forgotten process i.e. that it decays
into a ground state atom $a$ in mode $n$ and a photon $\varphi$ in any mode.
The decay due to the atoms $a$ in mode $n$, towards a special given mode of $
\varphi$, has a probability per unit time $\gamma \times \left[ a \right]
_{n}V$ (see the property (iv) of section 4 above). Therefore, the
probability of a decay towards the various modes $k$ of $\varphi$ which have
the same energy is

\begin{equation}
W_{f}=d_{\varphi}\times \gamma \times \left[a \right]_{n}V = \frac{\pi c^3}{
d_a \omega_\varphi \nu_\varphi}\frac{1}{t_{sp}}f_\varphi(\omega_\varphi -
\omega_{\varphi 0}) \times \frac{u_n(\omega_a)}{\hbar \omega_a}  \label{Wf}
\end{equation}
where $u_n(\omega_a) = [a]_n \times \hbar \omega_a$ is the energy density of
the ground state atoms in mode $n$.

At thermal equilibrium we have $[a]_n \ll [\varphi]_k$ (from Eq.~(\ref
{equilat})) which, using Eqs.~(\ref{Wem}) and (\ref{Wf}), implies $W_f \ll
W_{(em)}$, i.e. the probability for an excited atom to decay via stimulation
by a matter wave is much smaller than via stimulation by an electromagnetic
wave, and the "forgotten" process is practically negligible. However, this
may not be the case far from equilibrium. For example, one can imagine a
situation where one particular mode $[a]_n$ is strongly populated. This
would lead to decay preferentially towards that mode via the "forgotten"
process, which in turn increases the population of that mode, and so on.
That kind of mechanism is the basis of the experimental proposal for an atom
laser in \cite{Spreeuw}.

\section{The Einstein and the "forgotten" coefficients}

Now, on one hand we consider that the frequency width of the electromagnetic
spectrum of the radiation in the cavity is much larger than the width of the
natural decay spectrum of the atom (broad band approximation). On
the other hand we model the spectrum of the radiation in the cavity by
monochromatic radiation with angular frequencies $\omega _{\varphi \ell }$
where $\ell $ is an integer and $\omega _{\varphi \ell +1}-\omega _{\varphi
\ell }=\delta \omega _{\varphi }.$ The function $f_{\varphi }(\omega
_{\varphi \ell }-\omega _{\varphi 0})$ has a maximum for $\omega _{\varphi
\ell }\simeq \omega _{0};$ it is negligible for $\left\vert \omega _{\varphi
\ell }-\omega _{\varphi 0}\right\vert >\Delta \omega $ with $\Delta \omega
\ll \omega _{\varphi 0}$, and it fulfills the condition $\int_{0}^{\infty
}f_{\varphi }(\omega _{\varphi \ell }-\omega _{\varphi 0})d\omega =1.$
Assuming $\delta \omega _{\varphi }\ll \Delta \omega $ we can write~:
\begin{equation}
\sum_{\ell }f_{\varphi }(\omega _{\varphi \ell }-\omega _{\varphi 0})\delta
\omega _{\varphi }\simeq 1\hspace{1cm}.  \label{integ}
\end{equation}

We consider a given initial state characterized by the number $N_{a\varphi }$
of atoms $a\varphi $ with angular frequency $\omega _{a\varphi },$ and the
number $N_{a}$ of atoms $a$ with angular frequency $\omega _{a}.$ We use the
preceding results to calculate the number $dN$ of photons which are emitted
or absorbed during a time interval $dt,$ through the various processes.

Let us give the final results before we outline the derivation. One finds

(i) for spontaneous emission : $dN_{sp}=A$ $N_{a\varphi }~dt=\frac{1}{t_{sp}}
N_{a\varphi }~dt,$

(ii) for absorption : $dN_{abs}=B_{\left( abs\right) }~\rho _{\varphi
}\left( \omega _{\varphi 0}\right) ~N_{a}dt$ with
\begin{equation}
B_{(abs)}=\frac{d_{a\varphi }}{d_{a}d_{\varphi }}\times \frac{c^{3}}{
4\pi\hbar \omega _{\varphi 0} \nu _{\varphi 0}^2}\frac{1}{t_{sp}}
\label{Babs}
\end{equation}

(iii) for emission stimulated by the photons :

$dN_{em}=B_{\left( em\right) }~\rho _{\varphi }\left( \omega _{\varphi
0}\right) ~N_{a\varphi }dt$ with
\begin{equation}
B_{(em)}=\frac{1}{d_{\varphi }}\times \frac{c^{3}}{4\pi\hbar \omega
_{\varphi 0} \nu _{\varphi 0}^2}\frac{1}{t_{sp}}  \label{Bem}
\end{equation}

(iv) for the forgotten process (emission stimulated by the atoms) :

$dN_{f}=C\rho _{a}\left( \omega _{a0}\right) N_{a\varphi }dt$ with

\begin{equation}
C=\frac{1}{d_{a}}\times \frac{c^{3}}{4\pi \hbar \omega _{a0}\nu _{a0}^{2}}
\frac{1}{t_{sp}}  \label{CC}
\end{equation}
where $\omega _{a0} = \omega _{a\varphi }-\omega _{\varphi 0}$.

$B_{(abs)}$ and $B_{(em)}$ are the well known Einstein coefficients but $C$
is a new one.

We now detail the derivation of the above expressions. Let us for instance
calculate $B_{(em)}.$ We assume that, at a given angular frequency $\omega
_{\varphi },$ the two electromagnetic polarizations display the same energy
density $u_{k}\left( \omega _{\varphi }\right) $. Therefore the number of
photons produced during $dt$ by the stimulated emission due to the
electromagnetic radiation at angular frequency $\omega _{\varphi }$ is $
d_{\varphi }~W_{(em)}~N_{a\varphi }~dt$ where $W_{(em)}$ is given by Eq.~(
\ref{Wem}). The stimulated emission due to the radiations at the various
frequencies $\omega _{\varphi \ell }$ results in the number of emitted
photons $dN_{\left( em\right) }={\sum_\ell}d_{\varphi }~W_{(em)}~N_{a\varphi
}~dt.$ The function $f_{\varphi }\left( \omega _{\varphi }-\omega _{\varphi
0}\right) $ is a quickly varying function of $\omega _{\varphi }$ while $
u_{k}\left( \omega _{\varphi \ell }\right) $ is slowly varying within the
framework of the broad band approximation. More precisely, we assume $
u_{k}\left( \omega _{\varphi \ell }\right) \simeq u_{k}\left( \omega
_{\varphi 0}\right) ,$ $\omega _{\varphi \ell }\simeq \omega _{\varphi 0}$
and $\nu _{\varphi \ell }\simeq \nu _{\varphi 0}$, for $\omega _{\varphi
0}-\Delta \omega <$ $\omega _{\varphi \ell }<\omega _{\varphi 0}+\Delta
\omega .$ Therefore we can use these expressions for $u_{k}\left( \omega
_{\varphi \ell }\right) ,$ $\omega _{\varphi \ell }$ and $\nu _{\varphi \ell
}$ to calculate $dN_{\left( em\right) }.$ We obtain

\begin{eqnarray}
dN_{\left( em\right) } &=&{\sum_\ell }d_{\varphi }\times
W_{(em)}~N_{a\varphi }dt  \nonumber \\
&=&{\sum_\ell }d_{\varphi }\times \frac{1}{d_{\varphi }}\frac{\pi c^{3}}{
\hbar \omega _{\varphi \ell }^{~2}\nu _{\varphi \ell }}\frac{1}{t_{sp}}
u_{k}\left( \omega _{\varphi \ell }\right) ~f_{\varphi }\left( \omega
_{\varphi \ell }-\omega _{\varphi 0}\right) ~N_{a\varphi }dt  \nonumber \\
&\simeq&\frac{1}{d_{\varphi }}\frac{\pi c^{3}}{\hbar \omega _{\varphi
0}^{~2}\nu _{\varphi 0}}\frac{1}{t_{sp}}d_{\varphi }u_{k}\left( \omega
_{\varphi 0}\right) {\sum_\ell }f_{\varphi }\left( \omega _{\varphi \ell
}-\omega _{\varphi 0}\right) ~N_{a\varphi }dt
\end{eqnarray}
The spectral energy density is $\frac{d_{\varphi }~u_{k}\left( \omega
_{\varphi 0}\right) }{\delta \omega _{\varphi }}=\tilde{\rho}_{\varphi
}\left( \omega _{\varphi 0}\right) $ where $\tilde{\rho}_{\varphi }$ is
defined by relation (\ref{rhorho}) then
\begin{equation}
dN_{\left( em\right) }=\frac{1}{d_{\varphi }}\frac{\pi c^{3}}{\hbar \omega
_{\varphi 0}^{~2}\nu _{\varphi 0}}\frac{1}{t_{sp}}\tilde{\rho}_{\varphi
}\left( \omega _{\varphi 0}\right) {\sum_\ell }f_{\varphi }\left( \omega
_{\varphi \ell }-\omega _{\varphi 0}\right) \delta \omega_\varphi
~N_{a\varphi }dt.
\end{equation}
Finally, using Eqs.~(\ref{integ}) and (\ref{rhorho}) we find
\begin{equation}
dN_{\left( em\right) }=B_{(em)}~\rho _{\varphi }\left( \omega _{\varphi
0}\right) ~N_{a\varphi }dt
\end{equation}
with
\begin{equation}
B_{(em)}=\frac{1}{d_{\varphi }}\times \frac{c^{3}}{4\pi\hbar \omega
_{\varphi 0} \nu _{\varphi 0}^2}\frac{1}{t_{sp}}.
\end{equation}
This expression for $B_{(em)}$ can be found in many textbooks such as \cite
{Yariv} where it is calculated in a way very similar to the one presented
above. It is also derived in \cite{Hilborn}.

Expression (\ref{Babs}) for $B_{\left( abs\right) }$ is obtained by the same
method and we will not detail it here.

To calculate $C$ we express $dN_{f}$ using expression (\ref{Wf}) for $W_{f}$
. In this expression, $\omega _{a}=\omega _{a\varphi }-\omega _{\varphi \ell
}$ is a function of $\omega _{\varphi \ell }$ because the mode $m,$ of the
initial state is given. We assume that the $d_{a}$ atomic modes at the same
frequency $\omega _{a}$ display the same energy density $u_{n}\left( \omega
_{a}\right)$ so

\begin{eqnarray}
dN_{f}&=&{\sum_\ell }d_{a}\times W_{f}~N_{a\varphi }~dt  \nonumber \\
&=& {\sum_\ell }\frac{1}{d_{a}}\frac{\pi c^{3}}{\omega _{\varphi \ell }\nu
_{\varphi \ell }}\frac{1}{t_{sp}}f_{\varphi }\left( \omega _{\varphi \ell
}-\omega _{\varphi 0}\right) \times \frac{d_{a}u_{n}\left( \omega
_{a}\right) }{\hbar \omega _{a}}~N_{a\varphi }~dt.
\end{eqnarray}
We assume that $\omega _{a}$ and $u_{n}\left( \omega _{a}\right) $ are two
slowly varying functions of $\omega _{\varphi \ell },$ then, with the
conditions that we have already assumed, we find
\begin{eqnarray}
dN_{f} &=&\frac{1}{d_{a}}\frac{\pi c^{3}}{\omega _{\varphi 0}\nu _{\varphi 0}
}\frac{1}{t_{sp}}\times \frac{d_{a}u_{n}\left( \omega _{a0}\right) }{ \hbar
\omega _{a0}}{\sum_\ell }f_{\varphi }\left( \omega _{\varphi \ell }-\omega
_{\varphi 0}\right) ~N_{a\varphi }~dt  \nonumber \\
&=&\frac{1}{d_{a}}\frac{\pi c^{3}}{\omega _{\varphi 0}\nu _{\varphi 0}}
\frac{1}{t_{sp}}\times \frac{1}{\hbar \omega _{a0}}\frac{d_{a}u_{n}\left(
\omega _{a0}\right) }{\delta \omega _{\varphi }}~N_{a\varphi }~dt.
\label{dNf}
\end{eqnarray}
It is not obvious whether to introduce the spectral density $\rho _{a}\left(
\omega _{a0}\right) $ of the atoms because $\delta \omega _{\varphi }$ which
appears in the expression above is not the angular frequency separation
between two neighboring atomic rays.

In the bandwidth $\delta \omega _{\varphi }$, around $\omega _{a0},$ the
number of resonant atomic angular frequencies is given by expression (\ref
{dNa}):~$\delta \mathcal{N}_{a}=\frac{\omega _{a0}\nu _{a0}}{\pi c^{3}}
\delta \omega _{\varphi }\times V$. In the same bandwidth there is only one
resonant frequency for the photons (the consequence of the definition of $
\delta \omega _{\varphi }$) so $\delta \mathcal{N}_{\varphi }=\frac{\omega
_{\varphi 0}\nu _{\varphi 0}}{\pi c^{3}}\delta \omega _{\varphi }\times V=1$
(c.f. (\ref{dNphi})). Therefore $\delta \mathcal{N}_{a}=\frac{\omega
_{a0}\nu _{a0}}{\omega _{\varphi 0}\nu _{\varphi 0}}.$ The spectral energy
density is $\tilde{\rho}_{a}\left( \omega _{a0},T \right) =\frac{
d_{a}u_{n}\left( \omega _{a0}\right) \delta \mathcal{N}_{a}}{\delta \omega
_{\varphi }}=\frac{d_{a}u_{n}\left( \omega _{a0}\right) }{\delta \omega
_{\varphi }}\frac{\omega _{a0}\nu _{a0}}{\omega _{\varphi 0}\nu _{\varphi 0}}
.$ Substituting this into Eq.~(\ref{dNf}) and using the definition (\ref
{tilderhoa}) of $\tilde{\rho}_a$ we finally obtain
\begin{equation}
dN_{f} =C~\rho _{a}\left( \omega _{a0}\right) ~N_{a\varphi }~dt
\end{equation}
with
\begin{equation}
C=\frac{1}{d_{a}}\times \frac{c^{3}}{4\pi \hbar \omega _{a0}\nu _{a0}^{2}}
\frac{1}{t_{sp}}.
\end{equation}
One can notice the similarities between the expressions for $C$ and $
B_{(em)} $.

A permanent state is achieved when $dN_{sp}+dN_{em}+dN_{f}=dN_{abs}.$
Therefore, at thermal equilibrium, Eq.~(\ref{stat}) becomes
\begin{equation}
A\,N_{a\varphi }+B_{\left( em\right) }\rho _{\varphi}\left( \omega _{\varphi
0},T\right)\,N_{a\varphi}+C\,\rho _{a}\left( \omega
_{a0},T\right)\,N_{a\varphi }=B_{\left( abs\right) }\rho _{\varphi}\left(
\omega _{\varphi 0},T\right)\,N_{a}.  \label{all4}
\end{equation}
One can easily check that the expressions obtained above guarantee the
validity of Eq.~(\ref{all4}) at any temperature when $N_{a\varphi }$ and $
N_{a}$ satisfy Bose-Einstein statistics.

It is also possible to derive the conservation of energy and the
coefficients $B_{\left( em\right) },$ $B_{\left( abs\right) }$ and $C$
directly from Eq.~(\ref{all4}) where $N_{a\varphi }$ and $N_{a}$ fulfill
Bose-Einstein rather than the Maxwell-Boltzmann statistics. However, in this
paper, we chose to emphasize the elementary underlying processes providing,
for example, the transition probabilities $W_{(em)}$, $W_{(abs)}$ and $W_f$
(Eqs.~(\ref{Wem}), (\ref{Wabs}) and (\ref{Wf})) as functions of the observed
electromagnetic line-shape $f_\varphi(\omega_\varphi - \omega_{\varphi 0})$.

\section{Conclusion}

We have derived a general description of atom-photon interactions of the
form $a\varphi \rightleftharpoons a + \varphi$ that includes the "forgotten"
process of decay stimulated by the matter waves $a$, additionally to the
well known spontaneous decay and the decay stimulated by the electromagnetic
waves $\varphi$. Our aim was to provide a description based on fundamental
elementary principles, and to emphasize throughout this work the symmetry
between atomic and matter waves. To do so, we have followed standard
textbook descriptions of the involved known processes (in terms of
transition probabilities, energy densities, observed line shapes, and
Einstein coefficients) but applied them to also derive analogous expressions
for the "forgotten" process that cannot yet be found in textbooks.

Throughout this work the various mechanisms have been considered for the
case of thermal equilibrium. But, similarly to the well known photonic
processes, our results (in particular the transition probability $W_f$ and
coefficient $C$ of the "forgotten" process) are quite general and remain
valid far away from equilibrium. Non-equilibrium conditions (population
inversions etc.) are the fundamental ingredients of lasers, and one expects
the same to hold true for atom lasers i.e. sources of coherent matter waves.
Atom lasers based on matter wave amplification have not yet been built, but
several proposals for a practical realization based on the "forgotten"
process can be found in the literature \cite{XIIproc, Spreeuw, Borde}.
Indeed, any practical realization will necessarily involve conditions far
from thermal equilibrium, as can easily be seen from Eq.~(\ref{equilat}).
For example, a single atom $a$ inside a cavity corresponds to $[a]_n V = 1$
and therefore the thermal equilibrium condition (\ref{equilat}) is only
satisfied for temperatures $T \simeq 10^{13} K$ which are impossible to
attain in practice.

It is easy to generalize the results that we have obtained to the general
bosonic case $ab\rightleftharpoons a+b$ where $ab$, $a$ and $b$ are massive
or massless bosons of any kind (photons, atoms, molecules, etc.). If we
assume that the same concepts are valid for the decay of an atom and for the
chemical reaction $ab\rightleftharpoons a+b$, i.e. the concept of line shape
where $f_{\varphi }\left( \omega -\omega _{\varphi 0}\right) $ becomes $
f_{b}\left( \omega -\omega _{b0}\right) $ and the concept of stimulated and
spontaneous reactions, then all our considerations apply to the general
bosonic case. Moreover if we assume that in the same elementary angular
frequency interval $d\omega ,$ the number of resonance frequencies is much
higher for $b$ than for $a$, the developments of the preceeding sections
(sections 3 to 6) remain unchanged, except for the substitution $\varphi
\rightarrow b$ and $\nu_\varphi = \omega_\varphi/2\pi \rightarrow \nu_b =
\sqrt{(\hbar\omega_b)^2 - (M_b c^2)^2}/2\pi\hbar$ where $M_{b}$ is the mass
of $b$ in its fundamental state. Under these conditions, the formulae (\ref
{Wem}), (\ref{Wabs}), (\ref{Wf}), (\ref{Babs}), (\ref{Bem}), and (\ref{CC})
hold true for the general bosonic case, and the photon case is simply
recovered when setting $M_b = 0$ and $d_b = 2$. This simple generalization
further stresses the symmetry between electromagnetic and matter waves as
now the atom-photon interaction is just a special case of a general bosonic
interaction.

If a perfect symmetry exists between electromagnetic and matter waves, it
may be relevant to raise the question of the physical nature of holography
with atomic waves, a surely premature question for the time being.


\begin{thebibliography}{99}
\bibitem{Albert} Einstein A., {\it Zur Quantentheorie der Strahlung},
Phys. Zs. {\bf 18}, 121-128, (1917), first published in "Physikalische
Geselschaft Z\"{u}rich" Mitteilungen 18, 47-62, (1916), see also the
preliminary studies : {\it Strahlungs-Emission und Absorption nach den
Quantentheorie}" Deutsche physikalische Gesellschaft, Verhandlungen 18,
318-323 (1916).

\bibitem{XIIproc} Olshanii M., Castin Y., Dalibard J., {\it A model for an 
atom laser}, {\it Proc. XII Conf. on Laser Spectroscopy}, Inguscio M., 
Allegrini M., Sasso A., ed.,World Scientific, 7-12,(1996). 
Spreeuw R.J.C., et al., {\it Scheme for a bosonic atom laser}, 
{\it Proc. XII Conf. on Laser Spectroscopy}, Inguscio M., 
Allegrini M., Sasso A., ed.,World Scientific, 301-302,(1996). 
Bord\'e. Ch.J., {\it Amplification of atomic waves by 
stimulated emission of atoms}, 
{\it Proc. XII Conf. on Laser Spectroscopy}, Inguscio M., 
Allegrini M., Sasso A., ed.,World Scientific, 303-307,(1996).

\bibitem{Wiseman} Wiseman H.M., Collett M.J., {\it An atom laser based on
dark-state cooling} , Phys. Lett. {\bf A202}, 246-252, (1995).

\bibitem{Spreeuw} Spreeuw R.J.C., et al., {\it Laser-like scheme for
atomic-matter waves}, Europhys. Lett. {\bf 32}, 6, 469-474, (1995).

\bibitem{Borde} Bord\'{e}, Ch.J., {\it Amplification of atomic field by
stimulated emission of atoms}, Phys. Lett. {\bf A204}, 217-222, (1995).

\bibitem{Yariv} Yariv A., {\it Introduction to optical electronics},
Holt, Rinehart and Wilson, (1976).

\bibitem{Sargent} Sargent M. III, Scully M.O. and Lamb W.E. Jr, {\it 
Laser Physics}, Addison-Wesley, (1974).

\bibitem{Hilborn} Hilborn R.C., {\it Einstein coefficients, cross
sections, f values, dipole moments, and all that.},Am. J. Phys. {\bf 50},
11, 982-986, (1982) and references therein, (erratum in Am. J. Phys. {\bf 
51}, 5, 471, (1983)).

\bibitem{photon} The word "photon" was invented several years later but 
the concept of photon was already there.

\bibitem{Davisson} Davisson C.J. and Germer L.H., 
{\it Diffraction of electrons by a crystal of nickel},Phys. Rev. {\bf 30},
705-740, (1927).

\bibitem{Anderson} Anderson M.H., Enscher J.R., Matthews M.R., 
Wieman C.E. and Cornell E.A., {\it Observation of Bose-Einstein 
condensation in dilute atomic vapor},Science {\bf 269},
 198-201, (1995).

\bibitem{COW} Colella R., Overhauser A.W., Werner S.A., {\it Observation
of Gravitationally induced Quantum Interferences}, Phys. Rev. Lett. {\bf 
34}, 1472-1474, (1975).

\bibitem{Peters} Peters A., Chung K.Y., Chu S., {\it A measurement
ofgravitational acceleration by dropping atoms}, Nature {\bf 400},
849-852, (1999).

\bibitem{Snadden} Snadden M.J., et al., {\it Measurement of the Earth's
gravity gradient with an atom interferometer-based gtravity gradiometer},
Phys. Rev. Lett. {\bf 81}, 971-974, (1998).

\bibitem{Gustavson} L. Gustavson, A. Landragin, and M. A. Kasevich. {\it 
Rotation Sensing with aDual Atom-Interferometer Sagnac Gyroscope}. Class.
Quantum Gravity, 17:2385-98, (2000)

\bibitem{sensitivity} The sensitivity of atom interferometers 
is directly related to the number of
atoms (presently of order of $10^{5}$). Therefore, because of the Pauli
exclusion principle, the use of fermions is of less interest and not
considered here.

\bibitem{Landau} Landau L.D. and Lifschitz E.M., {\it Statistical Physics},
 Pergamon Press, New York, (1969).
 
 \bibitem{broad} The "broad band" approximation is valid when the radiation
  in the cavity
displays a line shape whose width is large compared to the width of the
natural lineshape of the atomic decay.

\bibitem{chronology}
The short chronology that we gave shows that this process could not be
considered before 1924. It was actually forgotten between 1924 and the 1990s
when it first appeared in the context of Bose-Einstein condensation \cite
{XIIproc,Wiseman,Spreeuw,Borde}.

\bibitem{rectangular} This is the case, for example, for an ideal 
rectangular cavity of suitable proportions.



\end{thebibliography}
\end{document}